\documentclass[twocolumn]{aa}
\usepackage{graphicx}
\usepackage{natbib}

\newcommand{\id}{\mbox{$\mathrm{d^{-1}}$}}

\newcommand{\kms}{\mbox{$\mathrm{km~s^{-1}}$}}
\newcommand{\Porb}{\mbox{$P_\mathrm{orb}$}}
\newcommand{\Psh}{\mbox{$P_\mathrm{sh}$}}

\newcommand{\Ion}[2]{#1{\,\scriptsize #2}}
\newcommand{\Ha}{\mbox{${\mathrm H\alpha}$}}
\newcommand{\Hb}{\mbox{${\mathrm H\beta}$}}
\newcommand{\Hg}{\mbox{${\mathrm H\gamma}$}}
\newcommand{\Hd}{\mbox{${\mathrm H\delta}$}}
\newcommand{\Twd}{\mbox{$T_{\mathrm{eff}}$}}
\newcommand{\Tdisc}{\mbox{$T_{\mathrm{disc}}$}}
\newcommand{\Rdisc}{\mbox{$R_{\mathrm{disc}}$}}
\newcommand{\Rwd}{\mbox{$R_1$}}
\newcommand{\Mwd}{\mbox{$M_1$}}
\newcommand{\Msec}{\mbox{$M_2$}}

\newcommand{\Msun}{\mbox{$\mathrm{M}_{\odot}$}}

\begin{document}
\title{Detection of the white dwarf and the secondary star in the new
SU\,UMa dwarf nova \object{HS\,2219+1824}
\thanks{Based in part 
on observations obtained at the German-Spanish Astronomical Center,
Calar Alto, operated by the Max-Planck-Institut f\"{u}r Astronomie,
Heidelberg, jointly with the Spanish National Commission for
Astronomy;
on observations made with the IAC80 and OGS telescopes, operated on
the island of Tenerife by the Instituto de Astrof\'\i sica de Canarias
(IAC) and the European Space Agency (ESA), respectively, in the
Spanish Observatorio del Teide of the IAC;
on observations made at the 1.2\,m telescope, located at Kryoneri
Korinthias, and owned by the National Observatory of Athens, Greece;
and on observations made with the William Herschel Telescope, which is
operated on the island of La Palma by the Isaac Newton Group in the
Spanish Observatorio del Roque de los Muchachos of the IAC.}}

\author{P. Rodr\'\i guez-Gil\inst{1} \and
        B. T. G\"ansicke\inst{1} \and
        H.-J. Hagen\inst{2} \and
        T. R. Marsh\inst{1} \and
	E. T. Harlaftis\inst{3} \and
	S. Kitsionas\inst{4} \and
	D. Engels\inst{2} 
	}

\offprints{P. Rodr\'\i guez-Gil,\\\email{Pablo.Rodriguez-Gil@warwick.ac.uk}}

\institute{
  Department of Physics, University of Warwick, Coventry CV4 7AL, UK
\and
  Hamburger Sternwarte, Universit\"at Hamburg, Gojenbergsweg 112,
  21029 Hamburg, Germany
\and
  Institute of Space Applications and Remote Sensing, National
  Observatory of Athens, P.O. Box 20048, Athens 11810, Greece
\and
  Institute of Astronomy and Astrophysics, National Observatory of Athens,
  P.O. Box 20048, Athens 11810, Greece 
}

   \date{Received 2004; accepted 2004}

\abstract{We report the discovery of a new, non-eclipsing SU\,UMa-type dwarf nova,
\object{HS\,2219+1824}. Photometry obtained in quiescence ($V\approx17.5$)
reveals a double-humped light curve from which we derive an orbital
period of $\simeq86.2$\,min. Additional photometry obtained during a
superoutburst reaching $V\simeq12.0$ clearly shows superhumps with a
period of $\simeq89.05$\,min. The optical spectrum contains
double-peaked Balmer and \Ion{He}{I} emission lines from the accretion
disc as well as broad absorption troughs of \Hb, \Hg, and \Hd\ from the white dwarf primary star. Modelling of the optical spectrum implies
a white dwarf temperature of
$13\,000\,\mathrm{K}\la\Twd\la17\,000\,\mathrm{K}$, a distance of
$180\,\mathrm{pc}\la d\la230\,\mathrm{pc}$, and suggests that the
spectral type of the donor star is later than M5. Phase-resolved
spectroscopy obtained during quiescence reveals a narrow \Ha\ emission
line component which has a radial velocity amplitude and phase
consistent with an origin on the secondary star, possibly on the
irradiated hemisphere facing the white dwarf. This constitutes the first detection of line emission from the secondary star in a quiescent \object{SU UMa} star.
%
}

\titlerunning{The new SU\,UMa star \object{HS\,2219+1824}}
\authorrunning{P. Rodr\'\i guez-Gil et al.}
\maketitle
%

\section{Introduction}
We are currently pursuing a large-scale survey for cataclysmic
variables (CVs), selecting candidates by the detection of Balmer
emission lines in the spectroscopic data from the Hamburg Quasar
Survey \citep{gaensickeetal02-2}. So far, this survey has proved to be
very efficient in identifying CVs that are relatively bright at
optical wavelengths but are characterised by either low-amplitude
variability, or by long outburst recurrence times, or by low X-ray
luminosities. Examples of systems discovered or independently
identified in our program include the deeply eclipsing, long-period
dwarf nova \object{GY\,Cnc} (\,=\,\object{HS\,0907+1902}, \citealt{gaensickeetal00-2}),
the rarely outbursting \object{SU\,UMa}-type dwarf nova \object{KV\,Dra} (\,=\,\object{HS\,1449+6415},
\citealt{nogamietal00-1}), the two intemediate polars
\object{1RXS\,J062518.2+733433} (\,=\,\object{HS\,0618+7336},
\citealt{araujo-betancoretal03-2}) and \object{DW\,Cnc} (\,=\,\object{HS\,0756+1624},
\citealt{rodriguez-giletal04-1}), the \object{SW\,Sex} stars \object{KUV\,03580+0614}
(\,=\,\object{HS\,0357+0614}, \citealt{szkodyetal01-1}) and \object{HS\,0728+6738}
\citep{rodriguez-giletal04-2}, and the old pre-CV \object{HS\,2237+8154}
\citep{gaensickeetal04-1}. Whereas all these systems belong to species
that are currently rare in the overall population of known CVs, their
intrinsic number may be comparable to, if not larger than, that of the
``classic'' CVs, which are either frequently in outburst, or X-ray
bright, or display obvious photometric variability. 

Here we report the discovery of a new short-period, SU\,UMa-type dwarf
nova with long recurrence time outbursts, \object{HS\,2219+1824}. This system belongs to the
relatively small group of dwarf novae whose optical spectra clearly
reveal the photospheric emission of their white dwarf primary stars. More
exceptional is, however, the fact that we detect in the quiescent
spectrum of \object{HS\,2219+1824} a narrow \Ha\ emission line component which we
believe to originate on the secondary star.

By inspection of the available sky surveys we discovered that the
USNO--A2.0 catalogue shows \object{HS\,2219+1824} in outburst with
$B=12.9$ and $R=13.6$. Both DSS\,1 and DSS\,2 show the system in
quiescence. ASAS-3 \citep{pojmanski01-1} monitored
\object{HS\,2219+1824} on 72 occasions between April 2003 and August
2004, and detected the July 2003 August superoutburst (also observed by us) as well as a second superoutburst reaching $V\simeq11.8$ on June 24 2004. Even
though sparse, the data collected so far shows that
\object{HS\,2219+1824} does not belong to the class of
\object{WZ\,Sge} dwarf novae with recurrence times of tens of
years. So far, no normal outburst of \object{HS\,2219+1824} has been
recorded.

We describe our observational data in Sect.\,\ref{s-observations}, the
analysis of the photometry and spectroscopy in Sect.\,\ref{s-ana_phot}
and Sect.\,\ref{s-ana_spect}, respectively, and discuss on the system parameters of \object{HS\,2219+1824} in
Sect.\,\ref{s-system_parameters}.

\section{\label{s-observations} Observations and data reduction}

\begin{table}[t]
\caption[]{\label{t-obslog}Log of observations.}
\setlength{\tabcolsep}{0.7ex}
\begin{flushleft}
\begin{tabular}[t]{lccccc}
\hline\noalign{\smallskip}
Date & UT & Filter/ & Mag. & Exp. & Frames \\
     &    & Grating               &     & (s)  & \\ 
\noalign{\smallskip}\hline
\noalign{\smallskip}
\multicolumn{5}{l}{\textbf{2.2\,m Calar Alto}}\\
2000 Sep 20 & 22:26 - 22:40 & $\mathrm{\displaystyle B-200\atop
  \displaystyle R-200}$ & $V\simeq17.3$ & 600 & 1/1 \\
\noalign{\smallskip}
\multicolumn{5}{l}{\textbf{1.2\,m Kryoneri Observatory}}\\
2002 Sep 17 & 00:36 - 03:16 & $R$ & $\simeq17.6$ & 90 &  97 \\
2002 Sep 17 & 22:40 - 23:11 & $R$ & $\simeq17.5$ & 60 &  25 \\
2002 Sep 18 & 19:04 - 23:41 & $R$ & $\simeq17.6$ & 60 & 210  \\
\noalign{\smallskip}
\multicolumn{5}{l}{\textbf{1\,m OGS}}\\
2003 Jul 08 & 03:27 - 05:24 & Clear & $\simeq15.7$ & 15 & 390 \\
2003 Jul 10 & 01:21 - 03:47 & Clear & $\simeq12.2$ & 15 & 515 \\
2003 Jul 11 & 02:47 - 04:06 & Clear & $\simeq12.2$ & 12 &  37 \\
2003 Jul 14 & 01:21 - 03:59 & Clear & $\simeq12.5$ & 12 & 548 \\
2003 Jul 16 & 02:20 - 05:31 & Clear & $\simeq12.8$ & 12 & 617 \\
\noalign{\smallskip}
\multicolumn{5}{l}{\textbf{0.82\,m IAC80}}\\
2003 Sep 25 & 00:19 - 03:14 & Clear & $\simeq17.4$ & 70  &  74  \\
2003 Sep 26 & 00:32 - 03:14 & Clear & $\simeq17.5$ & 70  & 119  \\
2003 Sep 27 & 23:31 - 03:10 & Clear & $\simeq17.5$ & 70  & 160 \\
2003 Sep 28 & 20:01 - 23:35 & Clear & $\simeq17.5$ & 70  & 147 \\
\noalign{\smallskip}
\multicolumn{5}{l}{\textbf{4.2\,m WHT}}\\
2003 Oct 19 & 19:35 - 22:58 & $\mathrm{\displaystyle R600B\atop
  \displaystyle R316R}$ & & 400 & 28 \\
\noalign{\smallskip}\hline
\end{tabular}
\end{flushleft}
\end{table}

\begin{figure}
\centerline{\includegraphics[width=7cm]{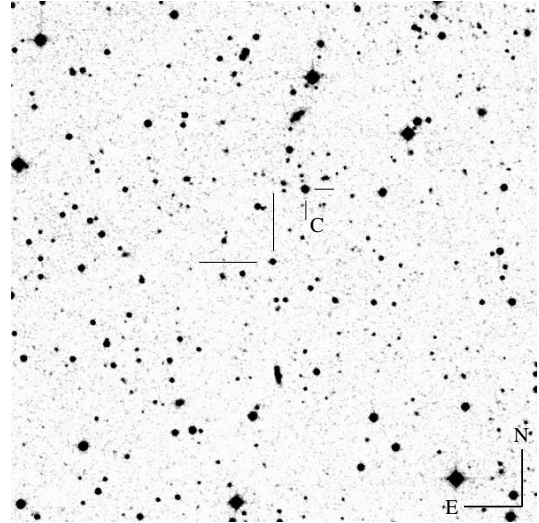}}
\caption[]{\label{fig-fc} $10\arcmin\times10\arcmin$ finding chart of
\object{HS\,2219+1824} obtained from the Digitized Sky Survey~2. The
coordinates of the CV are
$\alpha(2000)=22^\mathrm{h}21^\mathrm{m}44.8^\mathrm{s}$,
$\delta(2000)=+18\degr40\arcmin08.3\arcsec$. The star `C' has been
used as comparison for all the differential photometric data.}
\end{figure}

\subsection{Spectroscopy: Calar Alto}
A pair of blue/red identification spectra of \object{HS\,2219+1824} was
obtained at the 2.2\,m Calar Alto telescope with the CAFOS
spectrograph (Table\,\ref{t-obslog}). We used the B-200 and R-200
gratings in conjunction with a $2\arcsec$ slit, providing a spectral
resolution of $\simeq10$\,\AA~(FWHM). A standard reduction of these data was
carried out using the CAFOS \texttt{MIDAS} quicklook package. The
detection of strong Balmer emission lines confirmed the CV nature of
\object{HS\,2219+1824} and triggered the additional follow-up observations
described below. The CAFOS acquisition image obtained just before the
B-200/R-200 spectroscopy showed \object{HS\,2219+1824} at $V\simeq17.3$.

\subsection{Photometry: Kryoneri}
Differential $R$-band photometry of \object{HS\,2219+1824} was
obtained in 2002 September/October using the 1.2\,m Kryoneri telescope
equipped with a SI-502 $516\times516$ pixel$^2$ CCD camera
(Table\,\ref{t-obslog}). Aperture photometry was carried out on all
images using the \texttt{MIDAS} \& {\sc sextractor}
\citep{bertin+arnouts96-1} pipeline described by
\citet{gaensickeetal04-1}. The differential magnitudes of
\object{HS\,2219+1824} were derived relative to the comparison star
USNO--A2.0~1050--20233792, labelled `C' in Fig.\,\ref{fig-fc}. The
differential measurements were converted into $R$-band magnitudes
using the USNO--A2.0 magnitude of the comparison star,
$R_\mathrm{comp}=15.4$. The main source of uncertainty in this
conversion is the uncertainty in the USNO magnitudes, which is
typically $\simeq0.2$\,mag.

\subsection{\label{s-obs_ogs}Photometry: OGS}
The 1\,m Optical Ground Station (OGS) telescope at the Observatorio
del Teide on Tenerife was also used to perform time-resolved
photometry of \object{HS\,2219+1824}. The data were obtained between
2003 July 7 and 15 (Table~\ref{t-obslog}) using the Thomson
1024~$\times$~1024 pixel$^2$ CCD camera and no filter. Exposure times
of 12 and 15 seconds (depending on sky transparency) were adopted and
$2\times2$ binning and windowing were applied to improve the time
resolution. The raw images were de-biased and flat-field compensated,
and the instrumental magnitudes were obtained with the Point Spread
Function (PSF)-fitting packages within \texttt{IRAF}\footnote{\texttt{IRAF} is distributed by the National
Optical Astronomy Observatories, which is operated by the Association
of Universities for Research in Astronomy, Inc., under contract with
the National Science Foundation.}. Differential
light curves were then computed relative to the same comparison star
used for the Kryoneri data.

\begin{figure}
\centering
\includegraphics[angle=-90,width=9cm]{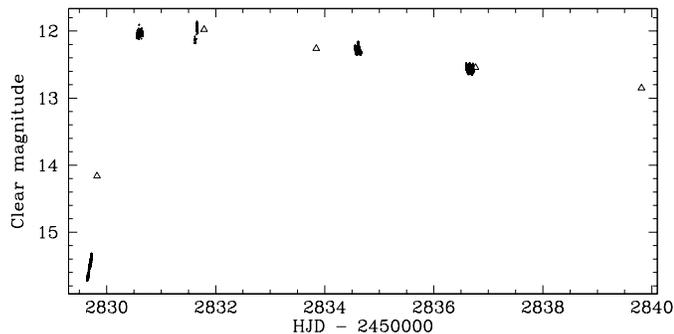}
\caption{\label{fig-ogslongterm} Superoutburst of \object{HS\,2219+1824} recorded
at the OGS telescope. Superhumps set in $\sim1$\,d after the outburst
maximum. The filterless magnitudes are approximatively equivalent to
$V$-band data. Shown as open triangles are the $V$ magnitudes of
\object{HS\,2219+1824} from the ASAS-3 archive.}
\end{figure}

The second night of observation at the OGS telescope offered us an
unusually bright \object{HS\,2219+1824}. Comparison with the
brightness measured on the previous night showed that the object had
brightened by nearly four magnitudes, indicating that an outburst was
in progress. The subsequent detection of superhumps (see
Fig.~\ref{fig-ogslongterm} and Sect\,\ref{s-ana_so}) confirmed that the
brightening was actually an \object{SU UMa}-like superoutburst.

The July 2003 superoutburst was also recorded by ASAS-3
\citep{pojmanski01-1}, and the $V$ magnitudes from the ASAS-3 archive
are plotted along with the OGS data in Fig.\,\ref{fig-ogslongterm}. 

\subsection{Photometry: IAC80}
Additional filterless photometry of \object{HS\,2219+1824} was obtained in
October 2003 with the 0.82\,m IAC80 telescope on Tenerife using the
Thomson 1024$\times$1024 pixel$^2$ CCD camera and an exposure time of
70\,s (Table\,\ref{t-obslog}). Only a small part of the CCD was read
out (in binning $2\times2$ mode) in order to improve the time
resolution. The data were reduced in the same fashion as described
above for the OGS (Sect.\,\ref{s-obs_ogs}). Differential magnitudes
of \object{HS\,2219+1824} were derived using the comparison star 'C'
(Fig.\,\ref{fig-fc}). The average magnitude of $\sim 17.5$\,mag
indicates that \object{HS\,2219+1824} was in quiescence during the 
IAC80 observations.

\begin{figure}
\centering
\includegraphics[angle=-90,width=9cm]{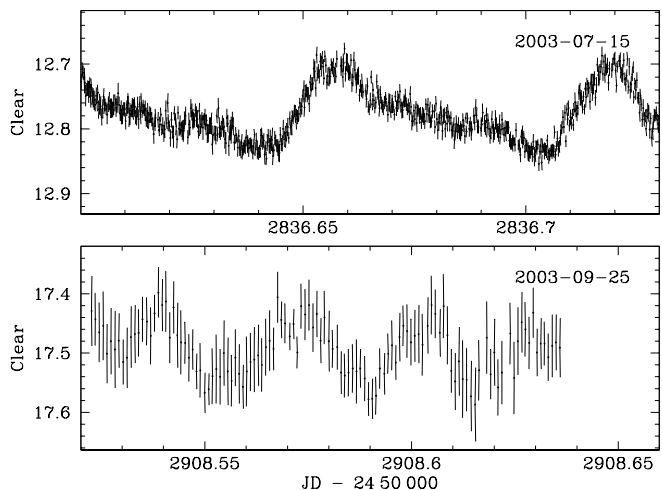}
\caption{\label{fig-lightcurves} Sample light curves of \object{HS\,2219+1824}
  obtained during the superoutburst (top panel) and in quiescence
  (bottom panel).}
\end{figure}

\subsection{Spectroscopy: William Herschel Telescope}
Time-resolved spectroscopy of
\object{HS\,2219+1824} in quiescence was performed on the night of 2003 October 19
with the 4.2\,m William Herschel Telescope (WHT) on La Palma and the
ISIS spectrograph. The blue arm was equipped with the R600B grating
and the $2048 \times 4100$ pixel$^2$ EEV12 CCD camera, while the R316R
grating and the $2047 \times 4611$ pixel$^2$ Marconi CCD were in place
on the red arm. A 1\arcsec~slit gave a spectral resolution of 1.8 and
3.2 \AA~(FWHM), respectively, covering the wavelength ranges
$\lambda\lambda4000-5000$ and $\lambda\lambda6100-9240$. The small
pixel size of both detectors produces oversampling of the best
possible resolution, so we applied a $1\times2$ binning (dispersion
direction) on both chips. This increased the signal in each wavelength
bin without losing spectral resolution.

A total of 28 blue and red spectra were obtained with an exposure time
of 400 seconds (see Table~\ref{t-obslog}). Spectra of the combined
light of Cu--Ar and Cu--Ne arc lamps were taken regularly in order to
achieve an optimal wavelength calibration. The raw images were
de-biased, flat-fielded and sky-subtracted in the standard way. The
target spectra were then optimally extracted \citep{horne86-1}. These
reduction processes were carried out within
\texttt{IRAF}. A third-order polynomial was fitted
to the arc wavelength--pixel function, the $rms$ being always less
than one tenth of the spectral dispersion on each arm.

\begin{figure*}
\includegraphics[angle=-90,width=18cm]{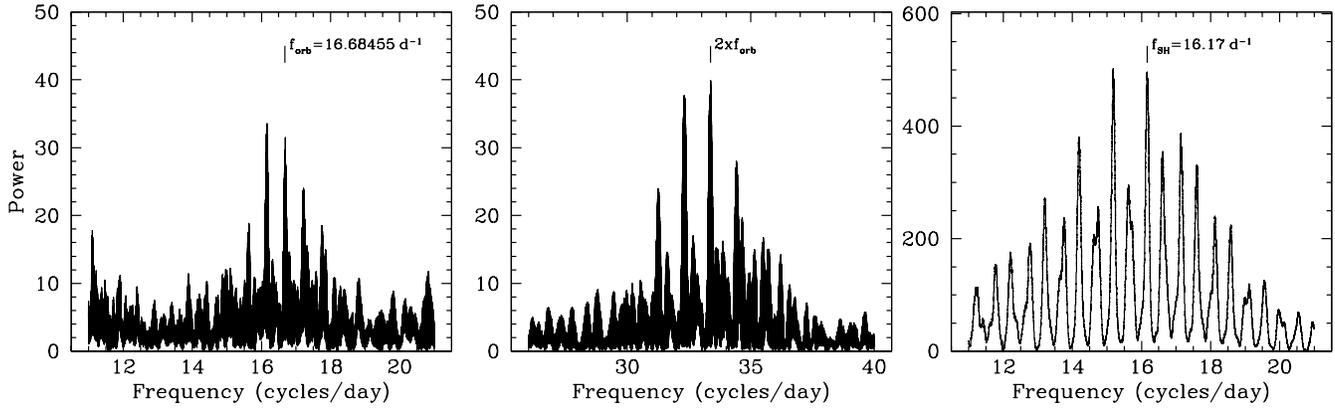}
\caption{\label{fig-aov_phot} Left and middle panels: AOV periodograms
  of the entire quiescent  data (Kryoneri \& IAC80). Right panel: AOV
  periodogram of the superoutburst data (OGS, 2003 July 11, 14 \& 16).}
\end{figure*}

\section{\label{s-ana_phot}Analysis: Photometry}

\begin{figure}
\centering
\includegraphics[angle=-90,width=9cm]{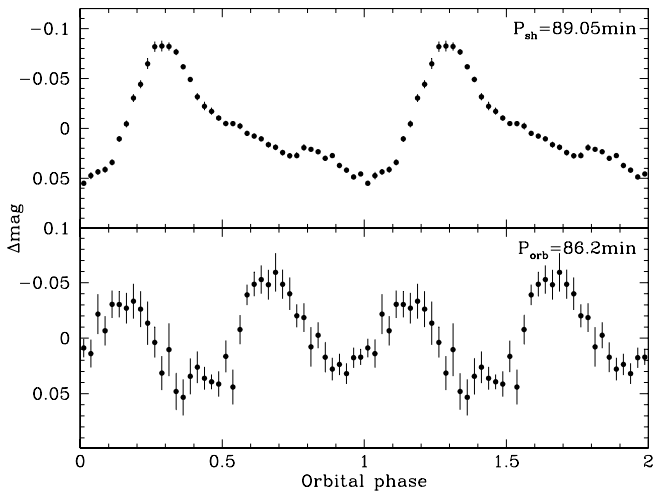}
\caption{\label{fig-folded_lc} Top panel: Superoutburst photometry
obtained on 2003 July 11, 14 \& 16 folded over the superhump period of
$\Psh=89.05$\,min and averaged into 40 phase bins. Bottom panel:
Quiescent photometry (Kryoneri 2002 and IAC80 2003) folded over the
orbital period of 86.2\,min and averaged into 40 phase bins. In both
cases, the mean magnitude of each individual night was subtracted
before folding and binning. The orbital cycle has been duplicated for continuity.}
\end{figure}

\subsection{\label{s-ana_qu}Quiescence}
During quiescence, the light curve of \object{HS\,2219+1824} displays a
hump-like structure with an amplitude of $\sim0.05$\,mag and a period
of $\sim40$\,min (bottom panel of Fig.\,\ref{fig-lightcurves}).  The
analysis-of-variance periodograms \citep[AOV,
][]{schwarzenberg-czerny89-1} computed from the Kryoneri and IAC80 quiescent data
(Fig.\,\ref{fig-aov_phot}) confirm this visual estimate, showing strong
signals at $\simeq43.1$\,min ($\simeq33.4$\,\id) and $\simeq44.6$\,min
($\simeq32.3$\,\id). Significant power is also found at twice these
periods. By analogy to a number of
short-period CVs which show quiescent orbital light curves dominated
by a double-humped structure (\object{WX\,Cet}: \citealt{mennickent94-1}; \object{WZ\,Sge}: \citealt{patterson98-1}; \object{RZ\,Leo}, \object{BC\,UMa}, \object{MM\,Hya}, \object{HV\,Vir}:
\citealt{pattersonetal03-1}; \object{HS\,2331+3905}:
\citealt{araujo-betancoretal04-1}), we identify the 43.1/44.6\,min
photometric periodicity with orbital variability, and hence $\Porb\simeq86.2$\,min
or $\Porb\simeq89.2$\,min. 

\subsection{\label{s-ana_so}The superoutburst}
Our July 2003 observations of \object{HS\,2219+1824} at the OGS caught the
system on the rise to outburst. The duration of the event and the
onset of superhumps clearly identify it as an \object{SU\,UMa}-type
superoutburst. We have computed an AOV periodogram from the OGS data
obtained on July 11, 14, and 16 (Fig.\,\ref{fig-aov_phot}). The
strongest signals are detected at $\simeq15.17$\,\id\
($\simeq94.92$\,min) and $\simeq16.17$\,\id\
($\simeq89.05$\,min). Both signals significantly differ from the two
possible orbital periods determined from the quiescent photometry. The longer period ($> \Porb$) and the shape of the outburst light curve are characteristic of a superhump wave, triggered by the precession of an eccentric accretion disc. 
The fractional period excess
$\epsilon=(\Psh-\Porb)/\Porb$ is $\epsilon=3.3$\% for
($\Porb=86.2$\,min, $\Psh=89.05$\,min), and $\epsilon=6.4$\% for
($\Porb=89.2$\,min, $\Psh=94.92$\,min). No dwarf nova with
$\epsilon>5$\% is known below the period gap
\citep{pattersonetal03-1}, which strongly suggests that $\Porb=86.2$\,min
and $\Psh=89.05$\,min are indeed the orbital and superhump periods of
\object{HS\,2219+1824}. In Fig.~\ref{fig-folded_lc} we show the superoutburst and quiescent light curves folded on 89.05 and 86.2 min, respectively.

\section{\label{s-ana_spect}Analysis: Spectroscopy}

\begin{figure}
\centering \includegraphics[angle=-90,width=9cm]{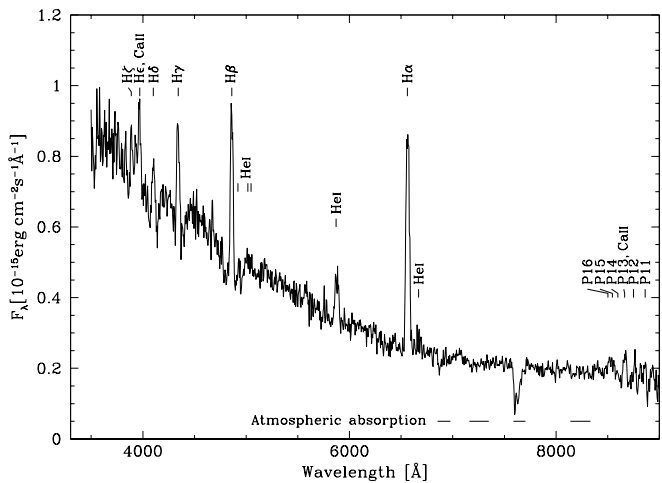}
\caption{\label{fig-idspec} Discovery spectrum of \object{HS\,2219+1824},
  obtained with the 2.2\,m Telescope at Calar Alto.}
\end{figure}

\subsection{\label{s-idspec} The optical spectrum of \object{HS\,2219+1824}}
The identification spectrum of \object{HS\,2219+1824}
(Fig.\,\ref{fig-idspec}) contains emission lines of neutral hydrogen
and helium. The Balmer lines show a relatively strong decrement and
\Hb\ to \Hg\ are embedded in very broad absorption troughs. Similar
broad Balmer absorption lines have been detected in a number of other
short period dwarf novae, e.g. \object{WZ\,Sge} \citep{greenstein57-1,
gillilandetal86-1}, \object{GW\,Lib} \citep{duerbeck+seitter87-1}, \object{BC\,UMa}
\citep{mukaietal90-1}, or \object{BW\,Scl} \citep{abbottetal97-1}.  In all
these systems the hypothetical identification of the observed Balmer
absorption as the photospheric spectrum of the white dwarf has been confirmed by
the unambiguous detection of the white dwarf at ultraviolet
wavelengths \citep{sionetal90-1, szkodyetal02-3, gaensickeetal04-2}.
The continuum displays a rather blue slope shortwards of \Ha, and a
nearly flat slope at the red end of our spectrum. No strong TiO
absorption lines, typical of mid-to-late M dwarfs are detected in the
red part of the spectrum.

\begin{figure}
\centering
\includegraphics[width=9cm]{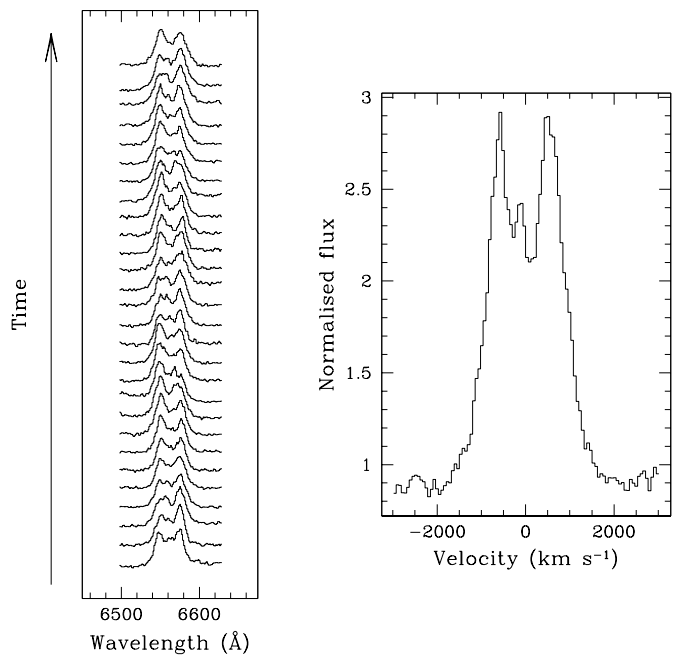}
\caption{\label{fig-profilevol} {\em Left panel}: evolution of the
\Ha~profile during the WHT/ISIS observations. The time co-ordinate
runs from UT 19:35 to UT 22:58 (see Table~\ref{t-obslog}). {\em Right
panel}: Selected \Ha~profile displaying three peaks.}
\end{figure}

The emission lines are clearly double-peaked in the higher resolution
WHT spectra (Fig.~\ref{fig-profilevol}), a clear indication of the presence
of an accretion disc in the system. Fig.~\ref{fig-profilevol}
illustrates the evolution of the \Ha~profile throughout our 2003
October 19 WHT/ISIS observations. The relative strength of the two
peaks clearly changes with time, possibly indicating the presence of
an emission S-wave moving inside the double peaks. Remarkably, many of
the profiles seem to have three peaks, revealing the presence of
another emission component in the lines (see right panel of
Fig.~\ref{fig-profilevol}).

\subsection{A model for the optical continuum}
Inspired by the similarity between the optical spectrum of
\object{HS\,2219+1824} and those of e.g. \object{WZ\,Sge}, \object{GW\,Lib}, \object{BC\,UMa} or \object{BW\,Scl}
(Sect.\,\ref{s-idspec}), which all reveal the spectra of
their white dwarf primaries, we have modelled the identification
spectrum of \object{HS\,2219+1824} with a simple three-component model
accounting for the emission of the white dwarf, the accretion disc,
and the secondary star. For the white dwarf, we use the model spectra
of \citet{gaensickeetal95-1}, and assume a surface gravity of $\log
g=8.0$ ($\simeq0.6$\,\Msun). The radius of the white dwarf is
determined from the \citet{hamada+salpeter61-1} mass-radius relation,
$\Rwd=8.7\times10^8$\,cm. Free parameters in the model are the white
dwarf effective temperature, \Twd, and the distance to
\object{HS\,2219+1824}. For the accretion disc, we use the isothermal/isobaric
hydrogen slab model described by \citet{gaensickeetal97-1,
gaensickeetal99-1}. Free parameters are the temperature of the slab
\Tdisc, the column density along the line of sight $\Sigma$, and a
flux scaling factor $\Omega$. For the secondary star, we use the
observed spectra of late-type main sequence stars from
\citet{beuermannetal98-1} as templates. We assume an equivalent
Roche-lobe radius of the secondary star of $R_{L_2}=9.5\times10^9$\,cm,
which corresponds to a mass ratio of $\Msec/\Mwd=0.14$  at the orbital
period of \object{HS\,2219+1824} (see the discussion in
Sect.\,\ref{s-system_parameters}). Free parameters for the secondary
star component are the spectral type Sp(2) and the distance to the
system.

\begin{figure}
\centering \includegraphics[angle=-90,width=9cm]{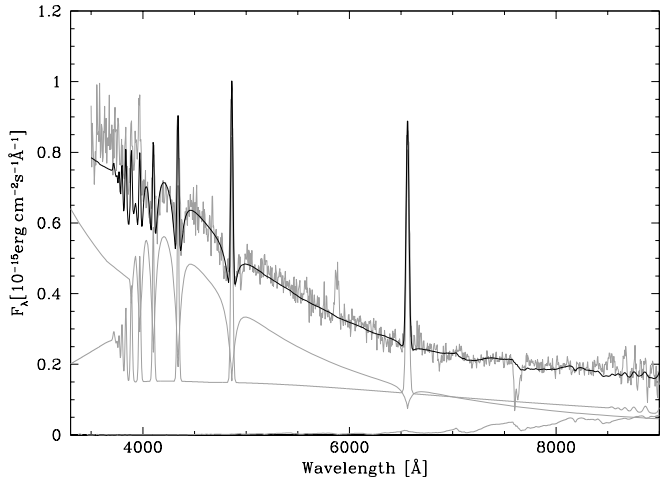}
\caption{\label{fig-optmodel} Three-component model of the optical
  spectrum of \object{HS\,2219+1824}. The observed spectrum is the topmost grey
  line. The grey lines below are a white dwarf model spectrum for
  $\Twd=16\,000$\,K, $\log g=8.0$, and $d=215$\,pc (broad absorption
  lines), the emission from an isothermal/isobaric slab with
  $\Tdisc=6000$\,K and $\Sigma=5.7\times10^{-2}\mathrm{g\,cm^{-2}}$,
  scaled to the observed \Ha\ flux (emission lines). The secondary
  star is represented by a M6 template, assuming
  $R_{L_2}=9.5\times10^9$\,cm and $d=215$\,pc. The sum of the three
  components is plotted as black line.}
\end{figure}

We started modelling of the observed optical spectrum of \object{HS\,2219+1824}
by matching the width and depth of the \Hb\ to \Hd\ absorption
profiles with a synthetic white white dwarf spectrum, and find
$13\,000\,\mathrm{K}\la\Twd\la17\,000$\,K and $180\,\mathrm{pc}\la
d\la230$\,pc~--~with the white dwarf contributing $\simeq65$\,\% at
5500\,\AA. In a second step, we add the emission of the isobaric slab,
scaling it to the observed \Ha\ flux. The main parameter determining
the Balmer decrement is \Tdisc, the optical depth ratio between the
emission lines and the continuum primarily depends on $\Sigma$. An
adequate match is found for $\Tdisc\simeq6000$\,K and
$\Sigma\simeq5.7\times10^{-2}\mathrm{g\,cm^{-2}}$. The ``disc''
temperature is the canonical temperature expected for low mass
transfer discs \citep{williams80-1}. For a distance of
$d\simeq200$\,pc, the radius of the ``disc'' implied by the flux
scaling factor is $\Rdisc\simeq2.3\times10^{10}$\,cm, well within the
Roche-lobe radius of the primary. This radius should, however, only be
considered as a rough estimate, as it is based on the assumption of our 
very simplistic model for the disc emission. Finally, we add the
spectral contribution of the secondary star. The observed flux at the
red end of the optical spectrum constrains Sp(2) to be later than M5
for a distance of $d\simeq200$\,pc. A model for $\Twd=16\,000$\,K, 
Sp(2)\,=\,M6.0, $d=215$\,pc and the disc parameters as above is shown
in Fig.\,\ref{fig-optmodel}.

\subsection{\label{s-rv1}Radial velocity curve analysis}

\begin{figure}
\centering
\includegraphics[width=9cm]{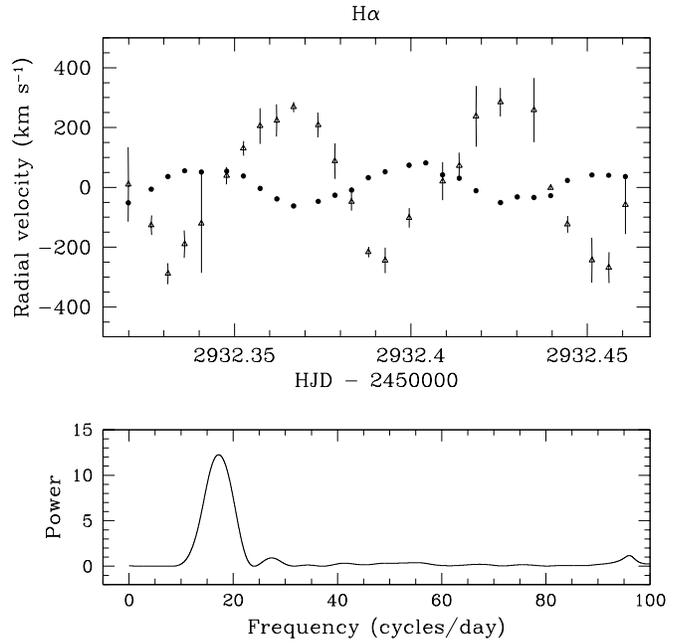}
\caption{\label{fig-ha_rvel} {\em Top}: \Ha~radial velocity curves of
\object{HS\,2219+1824} in quiescence computed from the line wings ({\em filled circles}) and the narrow component ({\em open triangles}). {\em Bottom}: Scargle
periodogram of the \Ha~wing velocities.}
\end{figure}

The light curves obtained during quiescence revealed a double-humped
variation at 86.2\,min (see Sect.~\ref{s-ana_qu}). In order to confirm
that this value represents the actual orbital period of
\object{HS\,2219+1824} we have analysed the radial velocity variations
of the \Ha~emission line, since this line is the least affected by the
broad absorption. Prior to measuring the velocities the individual
spectra were normalised using a low-order spline fit to the
continuum. The emission lines and the broad absorptions were masked
off the fitting procedure. We then resampled all the spectra on to an
uniform velocity scale centred on the \Ha~rest wavelength.  We
computed the radial velocity curves of \Ha~using the technique of
\cite{schneider+young80-2} for a Gaussian FWHM of 300 \kms and
different values of the Gaussian separation ($a=1200-2600$ \kms~in
steps of 100 \kms), following the technique of ``diagnostic diagrams''
\citep{shafter83-1, shafteretal95-1}. A sine function was fitted to
each curve to establish the $K_1$ velocity, and we find the maximum
useful separation to be $a \simeq 2300-2400$\,\kms, where $K_1\simeq
50$\,\kms. The \Ha~radial velocity curve obtained is shown in the top
panel of Fig.~\ref{fig-ha_rvel} (plotted as filled circles). A Scargle periodogram
\citep{scargle82-1} of the velocities produced a broad peak centred at
a period of $\sim 85$ min. A sine fit to the curve yielded a period of
$84.9 \pm 0.9$ min, which is consistent with the more accurate
photometric determination of $\Porb=86.2$ min.

For the time being, we follow the usual convention and assume that
high-velocity wings of the double-peaked emission lines originate in
the inner accretion disc, and track the orbital motion of the white
dwarf. A sine fit of the form $\gamma_1 -
K_1\,\sin\left[2\pi(t-T_{(1)0})/\Porb\right]$ to the \Ha~radial
velocity curve shown in Fig.~\ref{fig-ha_rvel} gives
$T_{(1)0}=2452932.3557 \pm 0.0003$ (HJD), $K_1=50 \pm 2$\,\kms~ and
$\gamma_1=8 \pm 1$\,\kms,with the orbital period fixed to 86.2 min. The
amplitude $K_1$ is quite typical for the radial velocity variation of
the emission line wings observed in short-period CVs
\citep[e.g.][]{thorstensen+fenton03-1}.  The lack of eclipses in
\object{HS\,2219+1824} does not allow us to establish absolute orbital
phases. However, under the assumption that the wings of the emission
lines come from disc material orbiting close to the white dwarf, the
red-to-blue crossing of the radial velocities provides a rough
estimate of the instant of inferior conjunction of the donor star
(i.e. zero phase).

\subsection{\label{s-rv2} Detection of chromospheric emission from the companion star}

\begin{figure}
\centering
\includegraphics[width=9cm]{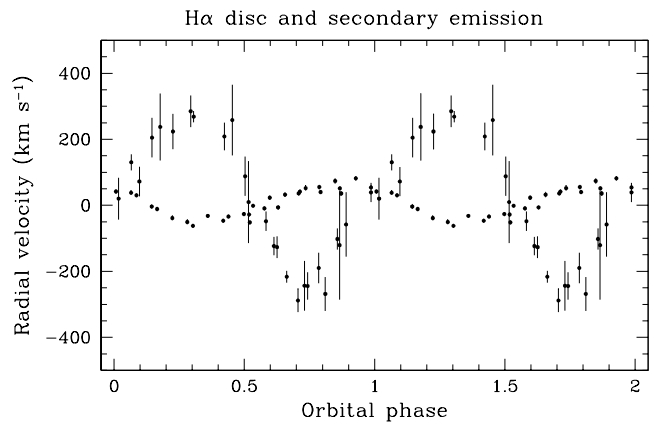}
\caption{\label{fig-bothrvcs} \Ha~radial velocity curves of the line
wings and the chromospheric emission from the companion. The
velocities are folded on the orbital period using $T_0=2452932.3485$
(HJD) and no phase binning has been applied. The orbital cycle has
been plotted twice for continuity.}
\end{figure}

The triple-peaked shape of the \Ha~profiles
(Fig.\,\ref{fig-profilevol}) suggests the presence of an additional
narrow emission line component. This feature is intense enough to
follow its motion by eye through almost all the individual
profiles. We measured the velocities of this component by fitting
single Gaussians after masking the rest of the line. The resulting
radial velocity curve is presented in Fig.~\ref{fig-ha_rvel} (plotted as open triangles).

A sine fit to the velocities of the narrow component with the orbital
period fixed provided $\gamma_2=28 \pm 7$ \kms~and $K_2=257 \pm 10$
\kms, but this time the curve is delayed by $\simeq 0.4$ orbital cycle
with respect to the assumed zero phase. If the phasing of the radial
velocity curve obtained from the \Ha~wings indeed reflects the
primary's motion, then the narrow emission originates at a location
offset by $\simeq140\degr$ in azimuth from the white dwarf. A very
likely source for a narrow emission line is the secondary star. The
velocity amplitude of the narrow component is much larger than that of
the broad line wings, in fact, $K_2=257$ \kms~is not unexpected for
the orbital motion of the low-mass companion in a short period CV. We
thus identify this narrow emission component as chromospheric emission
from the donor star, possibly due to irradiation from the white
dwarf/inner disc. Whereas narrow emission from the secondary star has
been identified in short period dwarf novae during outburst
\citep[e.g.][]{steeghsetal01-1}, \object{HS\,2219+1824} is the first
\object{SU\,UMa} dwarf nova where line emission from the secondary star is
unambiguously detected in quiescence\footnote{\cite{shafteretal84-1}
suggested the detection of the heated secondary during quiescence in
\object{IR\,Com} ($P_\mathrm{orb}=98.5$\,min) on the basis of a
perturbation in the radial velocity curves of the emission
lines. Secondary emission was also invoked in the eclipsing dwarf nova
\object{V893\,Sco} ($\Porb=109.4$\,min; \citealt{matsumotoetal00-1})
based on its \Ha~Doppler tomogram, but \object{V893 Sco} has never
undergone a superoutburst. The most reliable detection seems to have
been made in the long-period \object{SU UMa} star \object{TU Men}
($\Porb=168.8$\,min), whose \Ha~trailed spectra seem to show a narrow
emission moving with the phasing of the secondary
\citep{tappertetal03-1}, but its velocity semi-amplitude is almost the
same as the double peak semi-separation so one must be cautious when
interpreting their results.}.

We will adopt the new $T_{(2)0}=2452932.3485 \pm 0.0003$ (HJD) as the
time of inferior conjunction of the secondary. The radial velocity
curves of the disc and secondary emissions folded on the orbital
period are shown in Fig.~\ref{fig-bothrvcs}. The phase offset between
both velocity curves is not exactly 0.5 cycle, which is not surprising
as neither the velocity curves are perfectly sinusoidal nor are the
emissions expected to come from the centres of the stellar components
of the binary. Similar phase shifts have been observed in other
systems before \citep[e.g.][]{stoveretal81-1}.

\subsection{Trailed spectrograms and Doppler tomograms}

\begin{figure}
\centering \includegraphics[width=8.8cm]{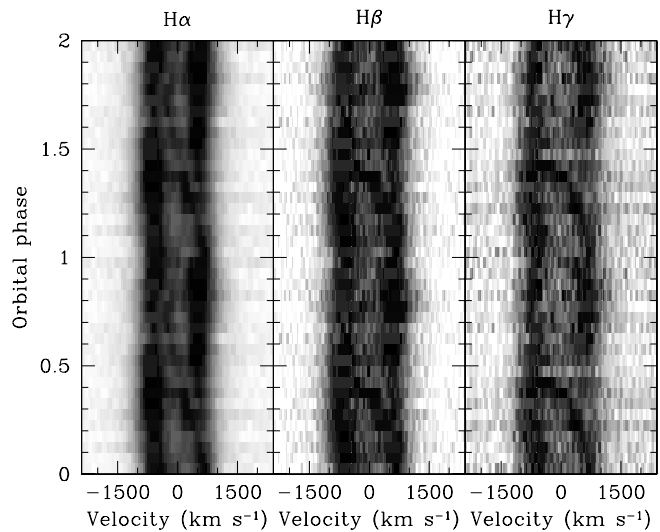}
\caption{\label{fig-trailed} \Ha, \Hb~and \Hg~trailed spectra
diagrams. The spectra have been folded on the orbital period and
averaged into 20 phase bins. Notice the narrow \Ha~emission component
from the secondary star. Another emission component with maximum blue
velocity at $\varphi \sim 0.6-0.7$ is evident in \Hb. Dark represents
emission and a full cycle has been repeated for continuity.}
\end{figure}

\begin{figure*}
\centering
\includegraphics[width=5.5cm,angle=-90]{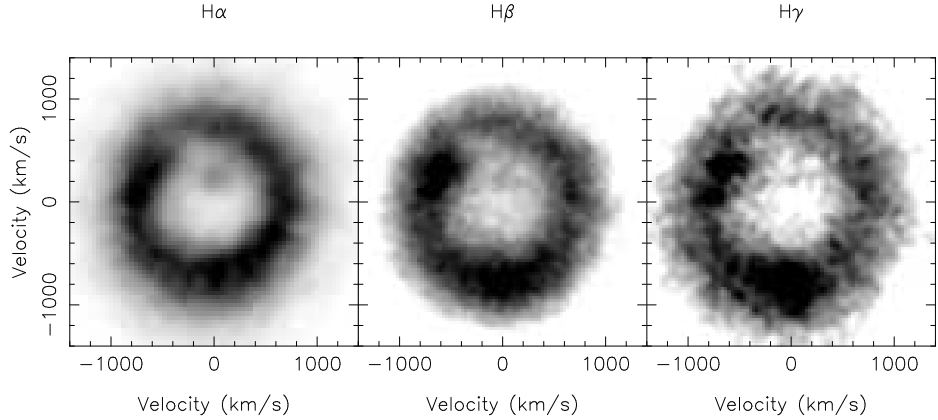}
\caption{\label{fig-doppler} \Ha, \Hb~and \Hg~Doppler tomograms. Dark
represents emission. Notice the \Ha~emission at the location of the
secondary and the two emission spots in \Hb~and \Hg.}
\end{figure*}

We have constructed trailed spectrograms of the \Ha, \Hb~and \Hg~lines
from the continuum-normalised data. In order to enhance the
signal-to-noise ratio we averaged the spectra into 20
orbital phase bins. The diagrams are presented in
Fig.~\ref{fig-trailed}. All the lines clearly show the alternating
changes in intensity of the double-peaked profiles. This is especially
obvious in \Ha, which also reveals the narrow
emission component from the irradiated secondary with maximum
excursion to the blue at $\varphi \sim 0.75$. Another emission S-wave
is visible in \Hb~moving in between the double peaks. This component
has the expected phasing for line emission from the bright spot
region, with maximum blue velocity at $\varphi \sim 0.6-0.7$, and is
strongest during the first half of the orbital cycle.

In order to map the line emission distribution in velocity space we
computed Doppler maps of \Ha, \Hb~and \Hg~using the maximum entropy
technique developed by \cite{marsh+horne88-1}. The tomograms are shown
in Fig.~\ref{fig-doppler}. All the maps show a ring of emission,
revealing the presence of an accretion disc. There is an emission spot
in the \Ha~map located at $(V_x\sim0,V_y\sim+200)$ \kms, where emission
from the secondary star is expected to be located in velocity
space. This strongly supports our hypothesis of an origin of the
narrow component seen in \Ha\ on the donor star in
\object{HS\,2219+1824}. Another emission spot is clearly seen in
\Hb~and \Hg, superimposed on the disc emission at
$(V_x\sim-450,V_y\sim+200)$ \kms. It is located close to the expected
position of the bright spot, as its phasing in the trailed spectra
already suggested. A third spot at $(V_x\sim 0,V_y\sim-600)$ \kms~can
be seen in the \Hb~and \Hg~tomograms, but the corresponding S-wave is
not evident in the trailed spectra diagrams.

\section{\label{s-system_parameters} System parameters}

\subsection{The mass ratio}
An estimate of the mass ratio $q=M_2/M_1$ of a CV can be obtained from
its superhump fractional period excess,
$\varepsilon=(P_\mathrm{sh}-P_\mathrm{orb})/P_\mathrm{orb}$, through
the relation \citep{patterson98-1}:
\begin{equation}
\varepsilon=\frac{0.23\,q}{1+0.27\,q}~.
\label{eq1}
\end{equation}
\noindent
In Sect.\,\ref{s-ana_so} we derived $\varepsilon=0.032$, corresponding
to a mass ratio of $q=0.14$. The uncertainty in $q$ is typically
dominated by the scatter in the observed $q(\varepsilon)$ relation (and
hence Patterson's \citeyear{patterson98-1} fit to this relation)
rather than by the uncertainty in $\varepsilon$. 

We have measured in Sect.\,\ref{s-rv1} and \ref{s-rv2} the amplitudes
of the radial velocity variations of the broad wings of the Balmer
lines and of the narrow emission line component detected in
\Ha. Attributing these radial velocity variations to material close to
the white dwarf and on the (irradiated face of the) secondary star,
respectively, implies $K_1\simeq50~\kms$ for the radial velocity
amplitude of the white dwarf and $K_2\simeq260$\,\kms\ for the
secondary, which gives a mass ratio of $q=K_1/K_2\simeq0.19$, somewhat
larger than the estimate from the fractional superhump period excess.
If, as we suggest, the chromospheric emission of the secondary star
originates primarily on the hemisphere facing the white dwarf then
our $K_2$ underestimates the true radial velocity amplitude of the
secondary, and this dynamical measure of $q$ is an upper limit. We
conclude that our observations suggest $0.14\la q\la0.19$.

\subsection{The size of the accretion disc}

We can also derive an estimate of the accretion disc radius from the
measured double peak velocity separation. We have measured the
semi-separation in the \Ha~profiles which are not contaminated by the
maximum velocity excursions of the S-waves, obtaining an average value
of $V_\mathrm{d} \approx 660$ \kms. This value gives a good estimate
of the projected velocity of the outer disc edge. \cite{smak81-1}
provided a relation between the Keplerian projected velocity of the
outer disc material, $V_\mathrm{K}(R_\mathrm{d})\,\sin i$, where $i$
is the orbital inclination of the system, and the velocity
semi-separation of the peaks, $V_\mathrm{d}$:
\begin{equation}
V_\mathrm{K}(R_\mathrm{d})\,\sin i \simeq 0.95\,V_\mathrm{d}~.
\label{eq2}
\end{equation} 
\noindent
Combining this expression with Kepler's Third Law, and using $q=0.14$,
$M_1\sim0.7$ \Msun, and $K_1 \la 50$ \kms, we find $R_\mathrm{d} \la
0.54\,R_\mathrm{L_1}$, where $R_\mathrm{L_1}$ is the distance from the
white dwarf to the inner Lagrangian point, given by
\cite{silber92-1}. Thus the accretion disc in \object{HS\,2219+1824}
in quiescence is probably quite small. The circularisation radius of
the system is given by:
\begin{equation}
R_\mathrm{circ} \approx \left(\frac{R_\mathrm{L_1}}{a}\right)^4 (1+q)~,
\label{eq3}
\end{equation} 
\noindent
where $a$ is the binary separation. For \object{HS\,2219+1824},
$R_\mathrm{circ} \approx 0.38\,R_\mathrm{L_1}$, smaller than the
derived disc radius. 

\section{Conclusions}
Based on our optical photometry and spectroscopy, we have identified
\object{HS\,2219+1824} to be an SU\,UMa-type dwarf nova with an orbital period
of $\simeq86.2$\,min and a superhump period of $\simeq89.05$\,min. The
superoutburst amplitude is $\sim5.5$\,magnitudes, reaching $V\simeq12$
at maximum. No normal outburst has been recorded so far. The broad
\Hb, \Hg, and \Hd\ absorption lines observed in the optical spectrum
of HS2219+1824 are consistent with the photospheric spectrum of a
$13\,000\,\mathrm{K}\la\Twd\la17\,000\,\mathrm{K}$ white dwarf at a
distance of $180\,\mathrm{pc}\la d\la230\,\mathrm{pc}$. The red part
of the optical spectrum constrains the spectral type of the donor to
be later than M5. Quiescent phase-resolved spectroscopy reveals an
unusual narrow \Ha\ emission line component which we tentatively attribute to
chromospheric emission from the irradiated inner hemisphere of the
secondary star. The superhump fractional period excess and the radial
velocities measured from this narrow \Ha\ component, as well as from
the broad wings of the double-peaked emission lines suggest a mass ratio of $0.14\la q\la0.19$.

\acknowledgements PRG and BTG were supported by a PPARC PDRA and a
PPARC Advanced Fellowship, respectively.  The HQS was supported by the
Deutsche Forschungsgemeinschaft through grants Re\,353/11 and
Re\,353/22. We are very grateful to Klaus Beuermann for giving us
access to his M-dwarf spectral templates, and to Patrick Schmeer for
alerting us of the 2004 outburst, as well as for drawing our attention
to the ASAS-3 program.

\bibliographystyle{aa}
\bibliography{aamnem99,aabib}
\end{document}